\documentclass[12pt,a4paper]{article}

\usepackage{graphicx,amssymb,amsmath}
\usepackage{cite}

\newcommand{\fsi}{\emph{fsi}~}

\begin{document}
\title{Physics at CELSIUS and COSY}

\author{H. Machner\thanks{Invited talk, HADRON 05, Rio de Janeiro} \\
Institut f\"{u}r Kernphysik \\ Forschungszentrum J\"{u}lich \\
52425 J\"{u}lich, Germany}

\date{\today}
\maketitle
\begin{abstract}
We review  some selected experimental results achieved at the
synchrotrons CELSIUS in Sweden and COSY in Germany. They concentrate
on meson production with emphasis on the underlying quark structure.
The project WASA at COSY is discussed and the search for symmetry
breaking in decays of $\eta$ and $\eta '$ mesons is highlighted.
\end{abstract}

\section{Introduction}\label{sec:Introduction}
CELSIUS at the The Svedberg Laboratory, Uppsala, Sweden and COSY at
the Research Center (FZ) J\"{u}lich, Germany have a lot of features in
common. Both have a cyclotron as injector, are synchrotrons with
beam cooling and operate as storage rings. On the other hand there
are differences. CELSIUS is slow ramping, has no external beams and
its circumference is much smaller than that of COSY, leaving only
space for two experiments. COSY is a rapid cycling machine with
presently four internal experiments installed and three external
target stations. Extraction of the beams is performed by stochastic
methods. The following text is organized into two parts. First we
will discuss some selected experimental results at both
accelerators. Then we will discuss in the second part the WASA
detector, which previously (summer 2005) operated at CELSIUS and is
presently disassembled and has been shipped to J\"{u}lich. It is
foreseen that the physics programme WASA at COSY will start in fall
2006.

\section{Selected Results}\label{sec:Selected-Results}
\subsection{OZI Rule Violation}\label{sub:OZI-Rule-Violation}
The Okubo-Zweig-Izuka (OZI) rule \cite{Okubo77} can be summarized in
simple terms that processes with continuing quark lines are favored
over those with discontinued quark lines. The $\omega$ is almost a
pure $u\bar u+d\bar d$ state while the $\phi$ is almost a pure
$s\bar s$ state. So production of the latter meson in a $pp$
reaction should be almost impossible. But quark mixing in these two
mesons makes the production of the $\phi$ possible. From the vector
mixing angle one gets $\sigma(pp\to pp\phi) / \sigma(pp\to pp\omega)
=\tan \alpha_v=0.004$. In $p\bar p$ annihilation the OZI rule was
found to be violated. More recently, DISTO measured the cross
sections for both reactions at the same beam momentum and found the
cross section ratio to strongly violate the OZI rule. However,
different phase space for both reactions introduce a deviation of
the cross section ratio from the OZU-rule. COSY TOF measured
$\sigma(pp\to pp\omega)$ cross sections and ANKE $\sigma(pp\to
pp\phi)$ cross sections. These data together with the DISTO data and
earlier SPES3 data for $\sigma(pp\to pp\omega)$ close to threshold
allows the estimation of the dependencies of the cross sections as
function of the excess energy $\epsilon$ which allows a comparison
that is equivalent to the same phase space.
\begin{figure}
\begin{center}
\includegraphics[width=6 cm]{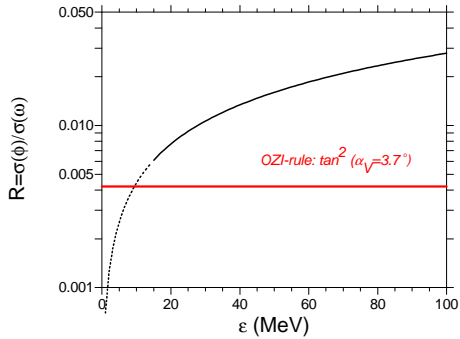}
\caption{The ratio of the two OZI rule related reactions as function
of the excess energy.} \label{fig:OZI}
\end{center}
\end{figure}
The data on $\phi$ production follow $s$-wave behavior while angular
distributions of $\omega$-production require several partial waves
to be fitted. We have therefore fitted a power law to the total
cross sections. The OZI-ratio is shown in Fig. \ref{fig:OZI}
together with the expectation value from the OZI-rule. The data show
a distinct deviation from it which is increasing with increasing
excess energy. There are several possible explanations why the
OZI-rule fails. One of them is the $s\bar s$ content in the proton.

\subsection{Associated Strangeness
Production}\label{sub:Associa-Strange-Production} Another topic
associated with $s\bar s$ production is its production in
\begin{equation}\label{equ:Lambda}
pp\to p\Lambda K^+
\end{equation}
and
\begin{equation}\label{equ:Sigma}
pp\to p\Sigma^0 K^+.
\end{equation}
\begin{figure}
\begin{center}
\includegraphics[width=8 cm]{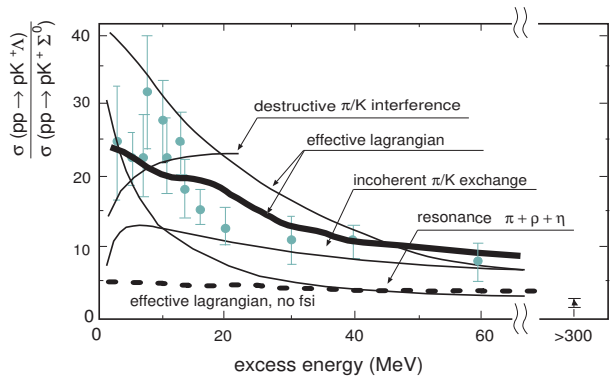}
\caption{The ratio of associated strangeness production as function
of the excess energy.} \label{fig:associated}
\end{center}
\end{figure}
These two reactions were measured by TOF at higher energies and by
COSY11 close to threshold. The ratio of both reactions is shown in
Fig. \ref{fig:associated}. It reaches a value 25--30 close to
threshold and decreases then to 8 at 60 MeV. This unexpected
behavior is studied within several models, including pion and kaon
exchange added coherently with destructive interference
\cite{Gasparian00} or incoherently~\cite{Sibirtsev99}, the
excitation of nucleon resonances~\cite{Shyam01,Shyam04} (labelled
effective Lagrangian), resonances with heavy meson exchange
($\pi,\rho,\eta$) \cite{Sibirtsev00} and heavy meson exchange
($\rho$, $\omega$ and $K^*$)~\cite{Shyam01,Shyam04}. The
corresponding curves are also shown in the figure. All models show a
decrease of the ratio with increasing excitation energy but none of
them accounts for all data except the one from Ref. \cite{Shyam04},
in which the sign of the poorly known coupling constant
$g_{pN(1650)}$ has been adjusted. It should be noted that $pY$-\fsi~
with $Y$ the hyperon is essential in reproducing the measurements.

This \fsi can be directly studied. HIRES at the Big Karl
spectrometer measures the excitation function at zero degree for
both reactions \ref{equ:Lambda} and \ref{equ:Sigma}. In TOF the
complete Dalitz plot is filled. A \fsi can be seen as enhancement
forming a line with constant mass of the $p\Lambda$ system. $N^*$
excitation is seen on the other hand as an enhancement as a line
with constant mass of the $K\Lambda$ system. An enhancement with
constant mass of the $pK^0$ system was interpreted by the TOF
collaboration as strange pentaquark $\Theta^+$ \cite{Abdel_Bary04}.
The group has in the meantime repeated the experiment in a longer
beam time and with an improved efficiency of their detector.
\subsection{A precision measurement of the $\eta$
mass}\label{sub:precisi-measure-eta-mass} Compared to other light
mesons, the mass of the $\eta$ is surprisingly poorly known. Though
the Particle Data Group (PDG) quotes a value of $m_{\eta}=547.75\pm
0.12~\textrm{MeV/c}^2$ in their 2004 review~\cite{PDG04}, this error
hides differences of up to 0.7~MeV/c$^2$ between the results of some
of the modern counter experiments quoted. The PDG average is in fact
dominated by the result of the CERN NA48 experiment,
$m_{\eta}=547.843\pm0.051~\textrm{MeV/c}^2$, which is based upon the
study of the kinematics of the six photons from the $3\pi^0$ decay
of 110~GeV $\eta$-mesons~\cite{Lai02}. In the other experiments
employing electronic detectors, which typically suggest a mass
$\approx 0.5~\textrm{MeV/c}^2$ lighter, the $\eta$ was produced much
closer to threshold and its mass primarily determined through a
missing-mass technique where, unlike the NA48 experiment, precise
knowledge of the beam momentum plays an essential part. GEM
performed a high precision determination of the $\eta$ meson mass.
The idea of the experiment is as follows. Three reactions were
measured at the same time at a beam momentum, where products of all
three reactions are detected simultaneously in the acceptance of the
Big Karl spectrometer. The reactions are
\begin{eqnarray}
p+d\to {^3H}+\pi^+ \label{eqn:t_pi} \\
p+d\to \pi^++{^3H} \label{eqn:pi_t} \\
p+d\to {^3He}+\eta \label{eqn:he_eta}.
\end{eqnarray}
The experiment simultaneously detected forward emitted pions and
backward emitted tritons in the c.m. system together with backward
emitted ${^3He}$ ions (doubly charged) as at a proton beam momentum
around 1640 MeV/c. Details of the experiment are found in Ref.
\cite{Abdel-Bary05}. The final result of this measurement is
\begin{equation}
m_{\eta}=547.311\pm 0.028\ \textrm{(stat.)} \pm 0.032\
\textrm{(syst.)\ MeV/c}^2\,.
\end{equation}

\begin{figure}
\begin{center}
\includegraphics[width=8cm]{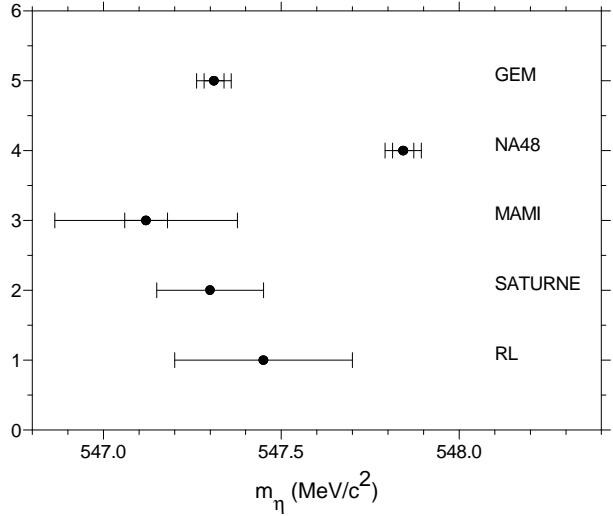}
\caption{The results of the $\eta$-mass measurements, in order of
publication date, taken from the Rutherford Laboratory
(RL)~\cite{Duane74}, SATURNE~\cite{Plouin92}, MAMI~\cite{Krusche95},
NA48~\cite{Lai02}, and GEM~\cite{Abdel-Bary05}. When two error bars
are shown, the smaller is statistical and the larger total.}
\label{eta_year}
\end{center}
\end{figure}

Our value of the mass of the $\eta$ meson is compared in
Fig.~\ref{eta_year} with the results of all other measurements
reported in the current PDG compilation~\cite{PDG04}. Though
significantly smaller than that reported in the NA48
experiment~\cite{Lai02}, it is in excellent agreement with the other
results.

\subsection{$\eta$ and $\eta '$ production in proton-proton
collisions}\label{sub:pp-production}

A comparison of neutral pseudoscalar meson production in $pp$
collisions should shed light on the reaction mechanism and the
interactions among the reaction partners. The $\eta$ and $\eta '$
mesons are the isospin zero partners of the $\pi^0$ which has
isospin one. The latter has a very weak interaction with nucleons
with respect to the nucleon-nucleon interaction. Furthermore, the
influence of an intermediate nucleon resonance, the $\Delta(1332)$,
was found only in the $Pp$ partial wave \cite{Betigeri02}.  In
contrast to this case, the $\eta$-nucleon and even more the
$\eta$-nucleus interaction is not that small.  The $N^*(1535)$ has a
strong coupling to the $\eta$-nucleon channel. On the other hand no
resonance is known to couple to the $\eta '$-nucleon channel. The
meson-nucleon interaction in all three cases can only be studied in
\fsi because of the short lifetime of the mesons.
\begin{figure}
\begin{center}
\includegraphics[width=6 cm]{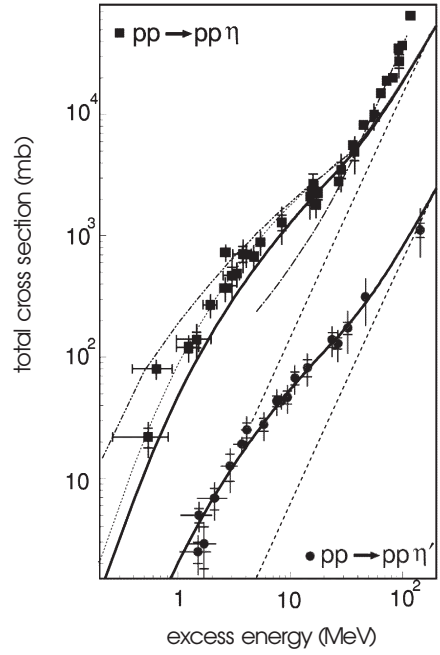}
\caption{Excitation functions of the total cross sections for $pp\to
\eta pp$ (circles) and $pp\to \eta 'pp$ reactions (squares). The
different curves are discussed in the text.} \label{fig:pp2eta}
\end{center}
\end{figure}
In Fig. \ref{fig:pp2eta} the total cross section for the reactions
as a function of the center-of-mass excess energy Q are shown. The
sources of data are given in Ref. \cite{Moskal04}. The dashed lines
indicate a phase-space integral normalized arbitrarily. The solid
lines show the phase-space distribution including the $^1S_0$
proton-proton strong and Coulomb interactions. In the case of the
$pp\to pp \eta$ reaction, the solid line was fitted to the data in
the excess energy range between 15 and 40 MeV. Additional inclusion
of the proton- $\eta$ \fsi is indicated by the dotted line. The
scattering length  and the effective range parameter have been
arbitrarily chosen. The dash-dotted line represents the energy
dependence taking into account the contribution from the $^3P_0\to
^1S_0s$, $^1S_0\to ^3P_0s$, and $^1D_2\to ^3P_2s$ transitions
\cite{Nakayama03}. Preliminary results for the $^3P_0\to ^1S_0s$
transition with the full treatment of the three-body effects are
shown as a dashed-double-dotted line \cite{Fixxx}. The absolute
scale of dashed-double-dotted line was fitted with an arbitrary
strength in order to demonstrate the energy dependence. While the
$\eta '$ production is nicely reproduced by phase space plus
$pp$-\fsi this is not the case for $\eta$ production. Most
importantly the Dalitz plot can not be explained by $p\eta$- and
$pp$-\fsi. The necessity for a rigorous three body calculation was
found \cite{Moskal04}.

\subsection{Proton-neutron final state
interaction}\label{sub:Proton--final-state-interaction} There has
been an extensive search for spin-singlet contribution in $pn$-\fsi.
Favorite reactions were $dp\to p \{pn\}$ and $dp\to \pi^+ \{pn\}$.
In both cases the pole (i. e. the deuteron) can also be measured.
There is a theorem due to F\"{a}ldt and Wilkin which connects the pole
to the continuum \cite{Faeldt97a}. Thus the absolute height of the
spin-triplet contribution in the continuum is given. The residual
cross section is usually attributed to the spin-singlet fraction,
however, the resolution and background conditions of most
experiments were not sufficient to unambiguously extract the
spin-singlet contribution. The characteristic feature of this
contribution is a very narrow peak due to an unbound pole at 60 keV,
which under these unfavorable conditions could not be seen directly.
Therefore, at COSY and CELSIUS the $\pi^+$ and proton were detected
in coincidence~\cite{Betsch99, Abaev01}. While identifying the
continuum channel well one loses the relative normalization with the
$\pi^+d$ final state. Therefore, one had to rely on Monte Carlo
simulations \cite{Betsch99}.

At GEM they measured the pions from $pp$ interactions with extremely
high resolution due to a 2 mm thin liquid hydrogen target and a
stochastically extracted beam which was electron cooled at injection
energy \cite{Abdel-Bary05a}. A missing mass resolution of
$\sigma=97$ keV was achieved for the deuteron. In addition an almost
halo-free beam resulted in a very small background. The F\"{a}ldt and
Wilkin \fsi -theorem yielded only 50$\%$ of the yield in the $pn$
continuum.
\begin{figure}\centering
\includegraphics[width=7 cm]{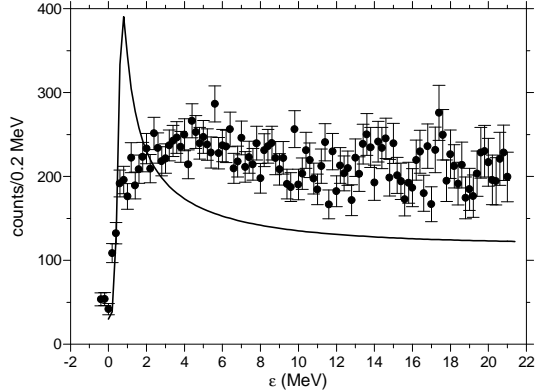}
\caption{Comparison of the measured $pn$ excitation energy spectrum
on a linear scale with the prediction the singlet cross section
folded with the present resolution. The error bars contain a tiny
contribution from the uncertainty in the acceptance correction.}
\label{fig:singlet}
\end{figure}
In Fig. \ref{fig:singlet}, a fit of the spin-singlet \fsi to the
data is shown. Obviously, that calculated cross section can not
account for the data. The reason for the discrepancy is not clear at
the moment. One explanation is that this is due to $D$-state effects
in the $pn$ system \cite{Abdel-Bary05a}. Another possibility might
be a failure of the F\"{a}ldt and Wilkin theorem which is exact only at
the pole position.

\section{Physics with WASA at COSY}\label{sec:WASA-COSY}
\subsection{Pseudoscalar meson
mixing}\label{sub:Pseudos-meson-mixing} The $QCD$ Hamiltonian can be
split into to parts
\begin{equation}
H_{QCD}  = H_0  + H_m
\end{equation}
with $H_0$ the Hamiltonian for massless quarks. In flavor SU(3) the
term containing the mass is given as
\begin{equation}
H_m  = \int {dx^3 \left( {m_u u\bar u + m_d d\bar d + m_s s\bar s}
\right)}.
\end{equation}
The latter term breaks chiral symmetry. The neutral pseudoscalar
mesons $\tilde{m}$ in ideal mixing are connected to the physical
mesons $m$ via
\begin{equation}
\left( {\begin{array}{*{20}c}
   {\tilde \pi }  \\
   {\tilde \eta }  \\
   {\tilde \eta '}  \\

 \end{array} } \right) = \left( {\begin{array}{*{20}c}
   {\frac{1}
{{\sqrt 2 }}\left( {u\bar u - d\bar d} \right)}  \\
   {\frac{1}
{{\sqrt 6 }}\left( {u\bar u + d\bar d - 2s\bar s} \right)}  \\
   {\frac{1}
{{\sqrt 3 }}\left( {u\bar u + d\bar d + s\bar s} \right)}  \\

 \end{array} } \right)\,\,\, = A\,\,\,\,\left( {\begin{array}{*{20}c}
   \pi   \\
   \eta   \\
   {\eta '}  \\

 \end{array} } \right).
\end{equation}
The matrix $A$ consists mainly of the $\pi^0-\eta$ mixing angle and
the $\eta -\eta '$ mixing angle. The former is given by
\begin{equation}
\sin \theta _{\pi \eta }    \equiv \frac{{\sqrt 3 }} {4}\frac{{m_d -
m_u }} {{m_s  - \hat m}}
\end{equation}
with $\hat m = (m_d  + m_u )/2$. A measurement of the $\pi^0-\eta$
mixing angle would therefore provide information about the up quark
and down quark mass difference.

A first attempt to measure this mixing angle was put forward by
Magiera and Machner \cite{Magiera00}. They proposed to measure the
ratio of backward emitted pions from the two reactions $pd\to
{^3He}\pi^0$ and $pd\to {^3H}\pi^+$ in the vicinity the $\eta$
threshold (below and above) in the reaction $pd\to {^3He}\eta$. They
argued that in the case of $\pi^0$ production this channel could be
enhanced below threshold due to $\pi^0-\eta$ mixing. Indeed an
effect was found in an experiment leading to
$\theta_{\pi\eta}=0.006\pm 0.005$ \cite{Abdel-Bary03}. However, Baru
et al. \cite{Baru03} claimed that a possible effect is not solely
due to the mixing but also $\eta -{^3He}$ should contribute. The
problem might be solved more cleanly by not measuring meson
production but studying meson decay instead. This can and will be be
done with the WASA detector at COSY.

\subsection{The WASA detector}\label{sub:WASA-detector}
\begin{figure}\centering
\includegraphics[width=12 cm]{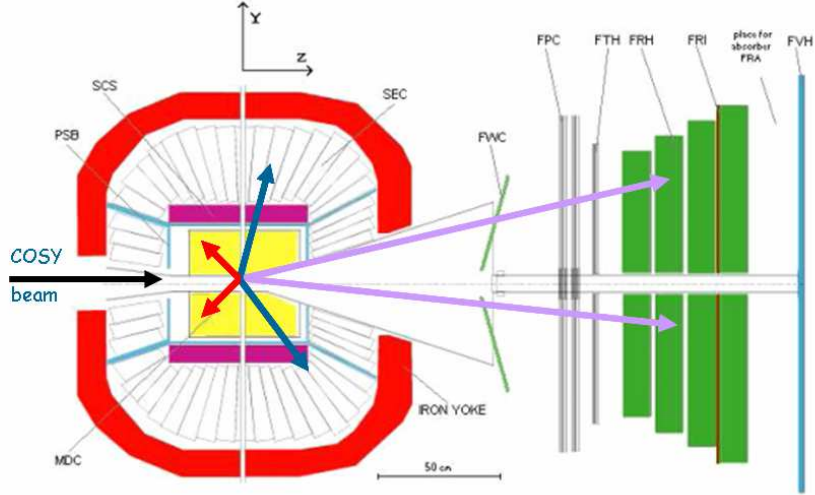}
\caption{Cross section to the WASA detector. The different
components are discussed in the text.}\label{fig:WASA}
\end{figure}

The WASA detector operated until summer of this year at the CELSIUS
facility. It is presently dissembled and has been shipped to J\"{u}lich
where it will be installed at COSY in late fall and spring of next
year. Fig. \ref{fig:WASA} shows a cross section through WASA.
WASA consists of a forward part (right) for measurements of charged
target-recoil particles and scattered projectiles and a central part
(left) designed for measurements of the meson decay products. The
forward part consists of eleven planes of plastic scintillators and
of proportional counter drift tubes. The central part consists of an
electromagnetic calorimeter of CsI(Na) crystals surrounding a
superconductive solenoid. Inside of the solenoid a cylindrical
chamber of drift tubes and a barrel of plastic scintillators are
placed. The arrows indicate a typical reaction $pp\to pp\eta '$ with
a subsequent decay of the $\eta '$ into an $\eta$ and two charged
pions. Finally, the $\eta$ decays into two $\gamma$'s.The two
protons will be measured in the forward detector, the charged pions
in the volume of the magnetic field and the $\gamma$'s in the CsI
crystals.
\subsection{$\eta$ and $\eta '$
decays}\label{sub:eta_eta-decays} One possibility of studying the
meson mixing angles are the decays of $\eta$ and $\eta '$ mesons. We
will first concentrate on the isospin forbidden decays of the $\eta
'$ into three pions. Instead of measuring the decays alone it was
proposed by Gross, Treiman and Wilczek \cite{Gross79} to measure the
ratios of the forbidden decays to the allowed decays into $\eta$ and
two pions:
\begin{equation}
\begin{gathered}
R_{ch}  = \frac{{\Gamma \left( {\eta ' \to \pi ^0 \pi ^ +  \pi ^ -
} \right)}}
{{\Gamma \left( {\eta ' \to \eta \pi ^ +  \pi ^ -  } \right)}} = PS_{ch} \sin ^2 \theta _{\pi \eta }  \hfill \\
R_{neut}  = \frac{{\Gamma \left( {\eta ' \to \pi ^0 \pi ^0 \pi ^0 }
\right)}}
{{\Gamma \left( {\eta ' \to \eta \pi ^0 \pi ^0 } \right)}} = PS_{neut} \sin ^2 \theta _{\pi \eta }.  \hfill \\
\end{gathered}
\end{equation}
$PS$ denotes the ratio of the three body phase. From the branchings
given by the PDG \cite{PDG04} one estimates for the charged pions
$R_{ch}<0.11$ while Gross et al. give an estimate of $1.49\times
10^{-3}$. The upper limit means that no events for this isospin
forbidden decay have been observed so far. For the neutral channel
the numbers are $(7.4\pm 1.2)\times 10^{-3}$ while the theory
predicts $1.37\times 10^{-3}$. The experimental number is based on
two experiments with a  count rate of $\approx 150$ counts in total.
One can expect that with the high luminosity anticipated at COSY
more than an order of magnitude more events will be recorded in a
couple of weeks.

The same information can in principle also be gained from the study
of the decay $\eta \to  \pi^+\pi^-\pi^0$. However, here Coulomb
corrections and theory input, both with some uncertainties
\cite{Leutwyler96b, Leutwyler96a}, are necessary.

Another approach is the study of the Dalitz plot of the decay
$\eta\to 3 \pi$. Here we will concentrate on the decay into three
neutral pions. The amplitude can be written as
\begin{equation}
\left| {A(\eta  \to 3\pi ^0 )} \right|^2  = 1 + 2\alpha z
\end{equation}
with
$z=\rho^2/\rho^2_{max}$ the relative radial distance. The distance
to the center of the Dalitz plot is $\rho$. A recent measurement of
the angle at WASA/CELSIUS \cite{Hoistadt05} yielded a preliminary
value of $\alpha=-0.027 \pm 0.015 \text{ ( stat. )}\pm 0.010\text{ (
syst.)}$.
\begin{figure}\centering
\includegraphics[width=8 cm]{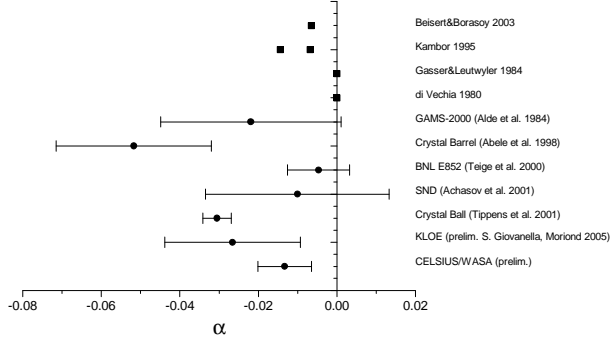}
\caption{The slope parameter $\alpha$ from various measurements and
calculations. The references are given in \cite{Tippens01} except
for the preliminary results.} \label{fig:alpha}
\end{figure}
This value is compared in Fig. \ref{fig:alpha} with previous
measurements and model calculations. The result is based on 37
thousand events. A much richer data sample can be expected at
WASA/COSY.

\begin{figure}[h]\centering
\includegraphics[width=13 cm]{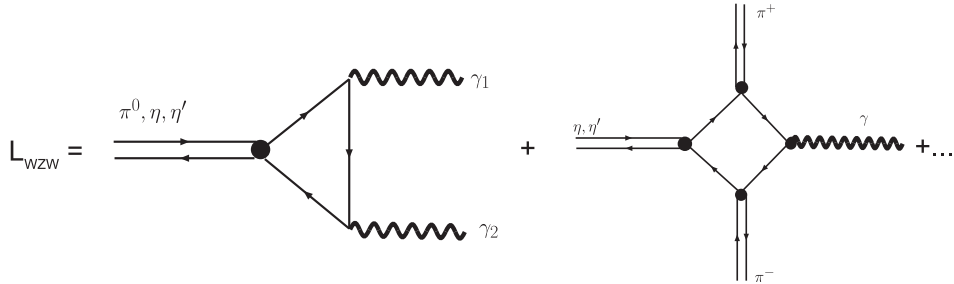}
\caption{The Wess-Zumino-Witten Lagrangian expanded in terms of the
triangle and box anomaly.} \label{fig:triangle_anomaly}
\end{figure}

In QCD with $N_f = 3$ there exists a non-Abelian anomaly which
breaks the chiral symmetry explicitly. In the effective chiral
Lagrangian this anomaly is appropriately reproduced by introducing
the Wess-Zumino action \cite{Wess-Zumino, Witten83}. The expansion
of the Wess-Zumino-Witten Lagrangian is shown in Fig.
\ref{fig:triangle_anomaly}.
Some interactions of the neutral pseudoscalar mesons have matrix
elements with the wrong parity. These are called anomalous
interactions. Among them are the two photon decay (triangle anomaly)
and the decay $\eta /\eta '\to \pi^+\pi^-\gamma$ (box anomaly).
These diagrams can be calculated within the model of hidden local
symmetry. Details can be found in a recent review \cite{Harada03}.
Benayoun et al. \cite{Benayoun03} constructed a set of equations
defining the amplitudes for $\eta /\eta '\to \pi^+\pi^-\gamma$  and
$\eta /\eta '\to 2\gamma$ at the chiral limit, as predicted from the
anomalous HLS Lagrangian and appropriately broken. For the decay
$\eta '\to \pi^+\pi^-\gamma$ they predict an invariant mass
distribution which is the same as the one from $\rho\pi^+\pi^-$
while for $\eta \to \pi^+\pi^-\gamma$ the center of gravity of the
distribution is shifted to $\approx 350$ MeV/c$^2$. In these
calculations the $\eta -\eta '$ mixing angle enters. The WASA
detector at COSY will allow one to measure the corresponding two
pion distribution with high statistics.

In the decay studies other rare decays will be measured. Most of
them violate $\mathcal{C}$-symmetry. The decay $\eta\to
\pi^+\pi^-e^+e^-$ violates $\mathcal{CP}$-symmetry outside the CKM
mechanism. Since this decay is flavor conserving, it is outside the
standard model. In a first run at WASA/CELSIUS already 25 candidates
were seen compared to 5 events in total in the literature.

\section{Acknowledgments} The author is grateful to B. Hoistadt, A.
Koukaz, P. Moskal, E. Roderburg, R. Schleichert and the members of
the GEM collaboration for their discussions concerning their data.
He is also thankful to R. Shyam for discussions about his
calculations and to J. B. Lieb for a critical reading of the
manuscript. The work was supported in part by Deutsche
Forschungsgemeinschaft MA 926/3-1.



\end{document}